
\magnification=\magstep1
\baselineskip  22 true   pt
\hsize 5 in\hoffset=.4 true in
\vsize 6.9 in\voffset=.4  true  in

\def\a{\alpha}
\def\b{\beta}
\def\g{\gamma}
\def\d{\delta}
\def\e{\epsilon}

\def\l{\lambda}
\def\om{\omega}

\def\m{\mu}
\def\n{\nu}
\def\r{\rho}
\def\s{\sigma}

\def\QNE{{F^a_{\mu\n}}}

\def\QNEB{{{\tilde{F}}^a_{\mu\n}}}

\def\QNII{{{\bar\eta}_{a\mu\n}}}

\def\lnrt{\longrightarrow}

\def\prm{\partial_M}
\def\prl{\partial_L}
\def\prmu{\partial_{\m}}
\def\prlam{\partial_{\l}}

\vfil
\hfil IP/BBSR/94-11\break

\hfil hep-th/9403085\break
\vfil
\centerline {\bf{Topological Electropoles in 4+1 Dimensional EYMCS Theory}}
\vskip 1.0 true cm
\centerline {{\bf{C.S.Aulakh}}\footnote\dag{
 Email:aulakh@iopb.ernet.in {\hskip 2 cm}\hfil {\hbox{{March 1994}}}}}
\centerline{ Institute of Physics, Sachivalya Marg}
\centerline{ Bhubaneshwar,751005,India}
\vskip 0.5 true cm
\centerline {and}
\centerline {\bf{V.Soni}}
\centerline{ National Physical Laboratory, Dr. K.Krishnan Marg}
\centerline{ New Delhi, India}
\vskip 2.0 cm
\centerline{\bf{ ABSTRACT}}

\noindent
A class of explicit exact solutions of Einstein Yang Mills Chern Simons
(EYMCS) theory corresponding to topological solitons carrying non-Abelian
topological electric charge is obtained. This verifies a conjecture made in
Ref.[1,2] regarding the stabilization of the corresponding charged
configurations in the theory without gravity .

\vfil
\eject

As is well known, the instanton solution${}^3$ of Yang-Mills theory
in 4 Euclidean dimensions can be viewed as a static topological
soliton of YM theory in
5 Minkowski dimensions. One also has (``caloron''${}^4$) solutions
when one dimension is compact (i.e on $M^4\times S^1$ ) and these
can serve${}^5$ as a paradigm for investigating the phenomenology of
solitons in theories with extra compact dimensions. Coupling the theory
to fermions  can induce (Abelian and
Non-Abelian) charges on the solitons due to the phenomenon of fermion
number fractionalization and its generalizations${}^6$.
The bosonic effective action which takes account of this phenomenon
involves the addition of the terms $\e^{MNLPQ} A_M F_{NL}^A F_{PQ}^A$
and $\om_5(A_M^A)$ (with
coefficents proportional to the number of fermions) to the YM action.
Here  $A_M,A_M^A$ are the Electromagnetic and YM gauge potentials
and $F_{MN}^A$ the YM field strength while $\om_5$ is the five
dimensional Chern-Simons term. These terms are also part of the $d=5$
Maxwell-Einstein-Yang-Mills supergravity action${}^{11}$ .

Thus the study of solutions of YMCS theory in $4+1 $ dimensions is well
motivated. In Ref.[5,1,2] we showed that the structure of the field equations
implies that the instantonic soliton , when trivially embedded in $SU(N)$
picks up a topological non-Abelian charge in the direction of the diagonal
generators of $SU(N)$ (other than isospin: thus for SU(3) the charge is in
the ``8'' direction ). This charge has an gauge covariant meaning in the
sense that the charged configuration transforms as a ${}^NC_2$ multiplet
(thus a triplet or antitriplet for $SU(3)$) of $SU(N)$.

Unfortunately the repulsive electrostatic energy of these configurations
makes them unstable towards growth of of the (arbitrary) scale parameter
($\r$) of the instanton configuration. As noted in Ref.[2] this is in
accord with the extension of Deser's theorem${}^7$ to the YMCS theory.
Recall that this theorem relies only upon the structure of the energy
momentum tensor to show that there are no regular finite energy static
solutions of the pure YM field equations which have non zero electric
field in five (non-compact) Minkowski dimensions. Since the CS term ,
being metric independent , does not contribute to the stress tensor
the proof  goes through unchanged. Thus, on $M^5$ , YMCS theory has
no charged solutions. On the other hand if one dimension is compact
the same arguments yield only a much weaker integral constraint .
Moreover the compactification of of one dimension implies that one component
of the gauge field behaves as a scalar in 4 dimensions and can give rise to
attractive forces that can counterbalance the electrostatic repulsion.
Another natural possibility is to couple the theory to gravity especially
since the Newtonian and Coulombic interactions have the same structure.
In fact following our conjecture in Ref.[2] we were able to show
numerically that if the Newtonian and Coulombic energies of the flat
 space solution were cancelled by tuning the couplings then EYMCS theory
(under a spherically symmetric 't Hooft ansatz) yielded essentially the flat
space (i.e self dual) solution for the YM fields. Deviation from this
fine tuned relation between the gauge and gravitational couplings resulted in
unacceptable (infinite energy) asymptotic behaviour. This is natural since a
net attraction or repulsion of a configuration with arbitrary scale should
drive the scale to zero or infinity. However the miracle of an exact force
balance for an extended object that is a solution of a highly nonlinear theory
requires more than the apocryphae of numerical analysis to inspire belief !
It is thus pleasing that using a method developed in Ref.[8] for a closely
related model (namely 5 dimensional EYM theories with Abelian CS tems plus the
$\e A GG$ coupling mentioned above ) we can obtain exact explicit
charged solitonic  solutions of the EYMCS theory. In this letter we give
solutions for $SU(3)$ EYMCS theory . Generalizations to $SU(N)$  and the
inclusion of electromagnetism are straightforward and will be reported
elsewhere.

We begin with the $SU(3)$ EYMCS action :
$$\eqalign {  S &= \int d^5 x\big(  {-1\over {16\pi G_5}} E R +
E ({1\over {2 g_5^2}} tr({F_{MN}} F^{MN}) +
 {{i N_f}\over {48\pi^2}} \omega_5\big) \cr
 \omega_5 &=\e^{MNLPQ}\quad tr(\prm A_N(\prl A_P A_Q + {3\over 2} A_L A_P A_Q)
 + {3\over 5} A_M A_N A_L A_P A_Q )}
\eqno(1)$$

Where $E= \sqrt{-det(g_{MN})}$, $g_{MN}$ is the five dimensional metric
and $G_5,g_5$ are the gravitational and gauge couplings in 5 dimensions with
mass dimensions -3 and -1/2 respectively.
Our conventions for gravitational quantities are those of Weinberg${}^9$ while
those for YM fields are as below :

$$\eqalign{ A_M &= A_M^A {{\l^A}\over 2i} \qquad\qquad tr \l^A\l^B =2\d^{AB}\cr
F_{MN} &=F_{MN}^A {{\l^A}\over 2i}=\partial_{[M} A_{N]} + [A_M,A_N]\cr
D_{M} &=\partial_{M}  + [A_M,\quad]\cr
A,B,...&= 1,...,8 ;\qquad M,N,...= 0,1,2,3,4\cr
\m,\n,...&= 1,2,3,4 ;\qquad a,b,...=1,2,3\cr}\eqno(2)$$

We choose a static metric of the Majumdar-Papapetrou${}^{12,13}$ form :

$$ ds^2 = -{1\over {(B^2)}} dt^2 + B dx^{\m}dx^{\m} \eqno(3) $$

here $dx^{\m}dx^{\m}$ is the Euclidean 4 metric which is appropriate${}^{13}$
when trying to find solutions in which the gravitational attraction and
Coulombic repulsion of a system of charges is in balance.
In Ref.[8] this form of the metric was also shown to be one appropriate
for finding solutions of $d=5$ supergravity theories that saturate a
generalized Bogomolny${}^{14}$ bound on the mass of charged
field configurations.

The field equations are ($\g=N_f/(48\pi^2)$):

$$\eqalign{ G_{MN} &= R_{MN} -{1\over 2} g_{MN} R = -8\pi G_5 T_{MN} \cr
     D_M(E F^{MNA}) &= - {{3\g g_5^2}\over 8} \e^{NMLPQ}\quad tr(\l^A
F_{ML} F_{PQ}) \cr} \eqno(4) $$

and $T_{MN}$ is the stress tensor:

$$ T_{MN} = {1\over {g_5^2}} (F^A_{ML} F_M^{A{}L} -
{1\over 4}g_{MN} F^A_{LP} F^{ALP}) \eqno(5)$$

The CS term gives rise${}^2$ to a source term
for the YM field equations which is proportional to the sum of the
 {\it{non-Abelian topological current}}:

$$\eqalign{T^{NA} &= {{3\g}\over 2}\e^{NMPLQ}\prm tr(\l^A (A_P\prl A_Q +
{2\over 3} A_P A_L A_Q))\cr
\prm T^{MA}&\equiv 0\cr}\eqno(6)$$

and the $SU(3)$ Noether current derived from the CS term.

We first recall the argument concerning charge induction${}^{1,2}$
for the reader's convenience . The constraint equation for static
configurations is :

$$D_{\m}(E D^{\m} A^{0A}) =-{{3\g g_5^2}\over 4}
tr (\l^A F_{\m\n} {\tilde F}_{\m\n})\eqno(7)$$

where ${\tilde F}_{\m\n} ={1\over 2}\e^{\m\n\l\s}F_{\l\s}$.
For a configuration with nonzero Pontryagin density
arising from static magnetic fields in the $SU(2)$ sector labelled by $a$.
it is clear that the rhs is non zero (only) for $A=8$. Thus it is inconsistent
to set $A_0^8=0$  for such static configurations. The minimum consistent
static ansatz is to take $\{A_{\m}^a,A_0^8\}$ to be nonzero. With
the metric (3)  eqn(7) becomes :

$$\prmu(B^2 \prmu A_0^8) = -({{{\sqrt 3} \g g_5^2}\over 8}
F^a_{\m\n} {\tilde F}^a_{\m\n})\eqno(8)$$

We now carry over the arguments of Ref.[8] to the present case :
For a {\it{flat space}} (anti) self -dual field configuration
(i.e $F^a_{\m\n}=\pm {\tilde F}^a_{\m\n}$)  the non-trivial space
components of the YM field equations reduce
(using the flat space Bianchi identity) to :

$$\prmu (\pm {1\over B} +
 {{{\sqrt 3} \g g_5^2}\over 2} A_0^8) {\tilde F}^a_{\m\n} =0 \eqno(9) $$

Thus if we choose ($k={2\over{{\sqrt 3}\g g_5^2}}$) :

$$A_0^8=\mp{k\over B}\eqno(10)$$

all the YM field equations are solved provided (8) is satisfied. Using
eqn.(10), eqn.(8) reads :

$$\prmu\prmu B=\mp{1\over{4 k^2}} F^a_{\m\n} {\tilde F}^a_{\m\n}\eqno(11)$$

It remains to show that if one fine tunes
the couplings then the Einstein equations also reduce to (11). First note that
due to flat space (anti) self-duality and the conformal
structure of the space sector of the metric the magnetic fields
$F^a_{\m\n}$ contribute {\it{only}} to $T_{00}$ :all other components
are identically zero . Thus the Einstein equations become
($\d=({{4\pi G_5 k^2}\over {g_5^2}} -{3\over 4})$) (no sum over $\l$):

$$\eqalign{ {3\over 2} {{\partial^2 B}\over B^4} +
\d ({{\prmu B}\over B})^2 &= \mp {{2\pi G_5}\over {g_5^2 B^4}}
F^a_{\m\n} {\tilde F}^a_{\m\n}\cr
\d \big(({{\prmu B}\over B})^2 -2 ({{\prlam B}\over B})^2\big) &=0 \cr
\d \prmu B\prlam B&=0 \cr}\eqno(12)$$

thus provided we choose $\d=0$ i.e
$$ {{16 \pi G_5 k^2}\over {3 g_5^2}}=1 \eqno(13a)$$

or, equivalently,

$$g_5^2 = 16 \pi ({{4\pi^2 G_5}\over{N_f^2}})^{1\over 3} \eqno(13b) $$

one finds that the Einstein equations reduce to the single equation (11) !
Thus ,like the electrostatic potential,  the static gravitational potential
B also has the Pontryagin density as its source. This is essentially
the reason for the possibility of a force balance .
Hence one only needs to solve one flatspace Poisson equation with a flatspace
(anti) selfdual configuration of nonzero Pontryagin index as its source
to obtain charged topological soliton solutions of EYMCS theory !
 Equation (13) is the force balance condition i.e it ensures that the
effects of gravitational attraction and electrostatic repulsion on the
self-dual configuration cancel exactly. Notice that spherical symmetry was
nowhere assumed.

It remains to solve for the potential function B. We shall restrict
ourselves here to completely regular solutions . Solutions with singularities
correspond to black holes with a specified charge to mass ratio superimposed
on the regular solutions found here and will be discussed elsewhere.
Consider a non-singular  (anti) self dual flat space YM field configuration
localized in a region V of extent R. Then the general solution of (11) is

$$ B =\a_1 \pm {1\over{16\pi^2 k^2}}
\int d^4y{1\over{(x-y)^2}} \QNE(y) \QNEB(y)
\eqno(14) $$

 From the regularity of the field strengths it is clear that B is everywhere
non singular . Moreover for $r>>R$ one has

$$ B= \a_1 + {{2\mid \n\mid}\over {k^2 r^2}} + O(1/r^3) \eqno(15) $$

where

$$\n={1\over{32\pi^2 }}
\int d^4x \QNE(x) \QNEB(x) \eqno(16) $$

is the flat space Pontryagin index of the configuration .

To identify the constant $\a_1$ and the mass and charge of the soliton
one defines as usual a new coordinate ${\tilde r}={\sqrt B} r$ so that
the metric becomes

$$ ds^2 =  -{1\over {B^2}} dt^2  +
(1-{{{\tilde r} B'({\tilde r})}\over {2 B}})^2 d{\tilde r}^2 +
{\tilde r}^2 d\Omega_3 \eqno(17) $$
where $d\Omega_3$ the metric on the unit 3-sphere.
For any localized system this metric should go over to the metric on $M^5$ as
${\tilde r}\lnrt \infty $. Then $B\lnrt 1$ and $r\lnrt {\tilde r}$.
Since one has  for a localized system in 5 dimensions${}^{10}$

$$g_{00} ={-1\over {B^2}}\lnrt -1 +{{8MG_5}\over {3\pi}}{1\over {\tilde r}^2}
+O({1\over {\tilde r}^3}) \eqno(18)$$

where M is the ADM mass , we immedeately get :

$$\a_1=1  \qquad\qquad M={{8\pi^2 \mid\n\mid}\over {g_5^2}} \eqno(19) $$

Furthermore the asymptotic form of the electric field :

$$ F^{0\m} = {{Q {\tilde x}^{\m}}\over{2\pi^2 {\tilde r}^4}} \eqno(20)$$

gives for the electrostatic charge $Q^8$ :

$$Q^8= -{{\n N_f g_5^2}\over {4 {\sqrt 3}}}\eqno(21)$$

If we restrict ourselves to (anti) self-dual configurations which can be
obtained via the 't Hooft${}^3$ ansatz :

$$A_{\m} =-\QNII \partial_{\n} ln\Pi(x^{\m}) \eqno(22)$$

we can obtain completely explicit solutions for all fields in terms of
the $x^{\m}$ coordinates. The self-duality condition reduces to solving
the equation $\Pi^{-1}\partial^2\Pi =0 $ for the ``superpotential''
$\Pi (x^{\m})$ .
The general solution  corresponding to N instantons with scale factors
$\r_i$ and locations $z_i^{\m}$ is :

$$ \Pi =1  +\sum_{i=1}^N {{\r_i^2}\over {(x-z_i)^2}}\eqno(23)$$

Although in this ansatz the gauge potentials are singular the
field strengths are everywhere regular and one can always transform to a
non-singular gauge. Note also that eqn.(11) is invariant under $SU(2)$
gauge transformations while $SU(3)$ gauge transformations give rise to the
$SU(3)$ multiplet structure described in Ref.[2].

To obtain the explicit form of the solution one need only remark that
for the solutions (23) one can write :

$$ (F^a_{\m\n})^2 = 2{\partial}^2 (\prlam Log(\Pi))^2 =
F^a_{\m\n}{\tilde F}^a_{\m\n} \eqno(24)$$

Thus the relevant solution of (11) is :

$$B=1  +\sum_{i=1}^N {{\b_i^2}\over {(x-{\tilde z}_i)^2}} -
{1\over{2 k^2}} (\prlam Log(\Pi))^2 \eqno(25)$$

The constants $\{\b_i, {\tilde z}_i^{\m}\}$ are fixed by demanding
regularity as $x^{\m}\lnrt z_i^{\m}$ :

$${\tilde z}_i^{\m}= z_i^{\m}\qquad\qquad \b_i={2\over {k^2}} \eqno(26)$$

Thus for instance for $N=1$ one has :

$$B= 1 + {{2 (r^2 + 2 \r^2)}\over {k^2 (r^2 +\r^2)^2}} \eqno(27) $$

Clearly B is everywhere regular and nonzero unless $\r=0$ so that there are
no horizons.

It is also easy to obtain explicit solutions when the asymptotic space is
$M^4 \times S^1$. One merely modifies the the metric (3) to ($m=1,2,3$ , R
is the radius of and $\theta$ the angle on $S^1$)

$$ds^2 = -{1\over {(B^2)}} dt^2 + B (dx^{m}dx^{m} + R^2 d\theta^2)\eqno(28)$$

and uses the periodic superpotential for N calorons.

In conclusion ,using the technique of Ref.[8], we have shown that
gravity can stabilize the solitonic
configurations of YMCS theory on $M^5$ and $M^4\times S^1$
which carry a topological non-abelian electric charge and
were introduced in Ref[1-3].
Our solutions suggest that the derivation of Bogomolny type bounds for
solitons in Einstein Maxwell theories (with non-minimal terms ) given in
Ref[8] may be extendable to a bound on the mass of solitons in terms
of the Casimir operator.
Our solutions are of interest in the context of Kaluza-Klein theories .
The program${}^{5,1,2}$ of building a simple paradigm
 for the phenomenology of
higher dimensional solitons by carrying out the collective quantization of
the global gauge zero modes of these solitons can now proceed.
\vskip.5 true cm
{\bf{Note Added}}

The expert reader will notice that our solution follows by
embedding the gauge group $SU(2)\times U(1)$ of Ref.[8] in $SU(3)$. However,
as explained in Ref.[2] the vacuum of the theory does {\it{not}} break
$SU(3)$ and the gauge covariance of the field equations implies that
one obtains solitons that are {\it{triplets}} of $SU(3)$. In this way
 an essentially non-Abelian structure is present.
The particular embedding chosen corresponds to a $SU(3)$ eigenstate in which
the isospin quantum number is zero.
The solutions of Ref.[8] are in turn closely related to those of Ref.[15].
A detailed comparison will be given elsewhere.

{\bf{Acknowledgements}}

We are grateful to Dieter Maison for useful discussions and participation
 in the intial stages of this work and for remarks leading to the
Note Added above.

\vfil
 \centerline {\bf{References }}

\item {1)} C.S.Aulakh ,Mod. Phys. Lett. {\bf{A7}},2469(1992).
\item {2)} C.S.Aulakh and V.Soni,Int. J.Mod.Phys.{\bf {A8}}1653(1993).
\item {3)} G.'t Hooft (unpublished); Phys. Rev.{\bf{D14}},3432(1976).
\item {} A.A.Belavin, A.M.Polyakov, A.S.Schwarz, Yu.S.Tyupkin,\hfil\break
Phys. Lett.{\bf{B59}},85(1975).
\item {} E.Corrigan and D.Fairlie, Phys. Lett.{\bf{B67}},69(1977).
\item {} R.Jackiw, C.Nohl. and C.Rebbi, Phys. Rev.{\bf{D15}},1642(1977).
\item {4)} B.J.Harrington and H.K.Shepard, Phys. Rev.{\bf{D17}},2122(1978).
\item {5)} C.S.Aulakh,Mod. Phys. Lett {\bf{A7}}2119(1992).
\item {} See also A.Strominger Nucl. Phys. {\bf{B343}} , 167(1990).
M.Kobayashi, Prog. Theor. Phys. {\bf{74}},1139(1985).
\item {6)} A.Niemi and G.W.Semenoff, Physics Reports {\bf{135}}, 99(1986).
and references therein ;Phys. Rev. Lett. {\bf {51}},  2077(1983).
\item{7)}S.Deser, Phys. Lett {\bf{B64}},463(1976).
\item{8)}G.W.Gibbons, D.Kastor, L.A.J.London,P.K.Townsend and\hfil\break
  J.Traschen, hep-th/9310118.
\item{9)} S.Weinberg, {\it{Gravitation and Cosmology}}, Wiley, New York(1972).
\item{10)}R.C.Myers and M.J.Perry, Ann. of Phys.{\bf{172}},304(1986).
\item{11)} M.G{\"u}nyadin, G.Sierra, and P.K.Townsend, Nucl. Phys.
 {\bf{B242}}, 244(1985); {\it{ibid}} {\bf{B253}}, 573(1985).
\item{12)} S.D.Majumdar, Phys.Rev.{\bf{72}},390(1947); A.Papapetrou, Proc. R.
Irish Acad. {\bf{A51}},191(1947).
\item{13)} R.C.Myers, Phys. Rev. {\bf{D35}},455(1987).
\item{14)} E.B.Bogomol'nyi, Sov. J. Nucl. Phys. {\bf{24}},449(1976).
\item{15)} A.Strominger, Nucl. Phys.{\bf{B343}},167(1990). For a review see
{\it{ Instantons and Solitons in Heterotic String Theory}}, Lectures given at
the Sixth Jorge Andre Sweica Summer School, PUPT-1278 (June 1991).
\vfil
\eject
\end